\documentclass[%
reprint,
superscriptaddress,
 amsmath,amssymb,
 aps,
]{revtex4-1}

\usepackage{graphicx}
\graphicspath{{figures/}}
\usepackage{amsmath,amssymb,amsfonts}
\usepackage{subfigure}
\usepackage[maxfloats=100]{morefloats}
\usepackage{dcolumn}
\usepackage{bm}
\usepackage{url}
\usepackage{algorithm}
\usepackage[noend]{algpseudocode}
\usepackage{mathtools}
\usepackage{comment}
\usepackage{multirow}
\usepackage{listings}
\usepackage{hyperref}
\usepackage{blindtext}

\usepackage{color}
\definecolor{rev1}{rgb}{0,0,0}

\makeatletter
\newenvironment{breakablealgorithm}
  {
   \begin{center}
     \refstepcounter{algorithm}
     \hrule height.8pt depth0pt \kern2pt
     \renewcommand{\caption}[2][\relax]{
       {\raggedright\textbf{\ALG@name~\thealgorithm} ##2\par}%
       \ifx\relax##1\relax 
         \addcontentsline{loa}{algorithm}{\protect\numberline{\thealgorithm}##2}%
       \else 
         \addcontentsline{loa}{algorithm}{\protect\numberline{\thealgorithm}##1}%
       \fi
       \kern2pt\hrule\kern2pt
     }
  }{
     \kern2pt\hrule\relax
   \end{center}
  }
\makeatother

\begin{document}



\title{An evolve-then-correct reduced order model for hidden fluid dynamics}

\author{Suraj Pawar}
\affiliation{%
 School of Mechanical and Aerospace Engineering, Oklahoma State University, Stillwater, OK 74078, USA\\
}%

\author{Shady E. Ahmed}
\affiliation{%
 School of Mechanical and Aerospace Engineering, Oklahoma State University, Stillwater, OK 74078, USA\\
}%

\author{O. San}
\email{osan@okstate.edu}%
\affiliation{%
 School of Mechanical and Aerospace Engineering, Oklahoma State University, Stillwater, OK 74078, USA\\
}%

\author{A. Rasheed}
\affiliation{Department of Engineering Cybernetics,\\ Norwegian University of Science and Technology, N-7465, Trondheim, Norway.\\
}%






\date{\today}

\begin{abstract}
In this paper, we put forth an evolve-then-correct reduced order modeling approach that combines intrusive and nonintrusive models to take hidden physical processes into account. Specifically, we split the underlying dynamics into known and unknown components. In the known part, we first utilize an intrusive Galerkin method projected on a set of basis functions obtained by proper orthogonal decomposition. We then formulate a recurrent neural network emulator based on the assumption that observed data is a manifestation of all relevant processes. We further enhance our approach by using an orthonormality conforming basis interpolation approach on a Grassmannian manifold to address off-design conditions. The proposed framework is illustrated here with the application of two-dimensional co-rotating vortex simulations under modeling uncertainty. The results demonstrate highly accurate predictions underlining the effectiveness of the evolve-then-correct approach toward realtime simulations, where the full process model is not known a priori.

\end{abstract}

\maketitle


\section{Introduction}
\label{sec:intro}

In fluid mechanics community, flow systems are often characterized by excessively large spatio-temporal scales. This puts a severe restriction on efficient deployment of practical applications which usually require near real-time, and many-query responses. Traditional full-order simulations cannot achieve this since their computational cost cannot be handled even with the largest supercomputers. Therefore, reduced order modeling emerges as a natural choice to tackle the computationally expensive problems with acceptable accuracy \cite{noack2003hierarchy, lucia2004reduced, quarteroni2014reduced,noack2011reduced,taira2017modal,benner2015survey, taira2019modal,puzyrev2019pyrom}. In reduced order models (ROMs), the evolution of the most important and relevant features is tracked rather than simulating each point in the flow field. These features represent the dynamics of the underlying flow patterns, also called basis functions or modes.

Proper orthogonal decomposition (POD) is a common technique to extract the dominant modes that contribute most to the total system's energy \cite{berkooz1993proper}. POD, coupled with Galerkin projection (GP), has been used for many years to formulate ROMs for dynamical systems \cite{ito1998reduced, rowley2004model,stankiewicz2008reduced, akhtar2009stability}. In these ROMs, the full-order set of equations is projected on the reduced space resulting in a dynamical system (in modal coefficients) with much lower order than the full order model (FOM). Clearly, this type of ROMs is limited to systems for which we have access to the governing equations. That is why they are called intrusive ROMs. However, in many situations, there is a discrepancy between the governing equations and the observed system. This might result from the approximation of underlying phenomena, incorrect parameterization, or insufficient information about source terms. This is, in particular, evident in atmospheric and geophysical systems where many complex phenomena and source terms interact with each other in different ways \cite{krasnopolsky2005new,krasnopolsky2006complex,brenowitz2018prognostic}.

Conversely, pure data-driven techniques, also known as nonintrusive, solely depend on observed data to model the system of interest. Therefore, more complicated processes can be modeled without the need to formulate them with mathematical expressions \cite{peherstorfer2016data}. Machine learning (ML) tools have shown substantial success in the fluid community identifying the underlying structures and mimicking their dynamics \cite{kutz2017deep, brunton2019machine, brenner2019perspective, duraisamy2019turbulence, lee2018model, xiao2019reduced}. However, end-to-end modeling with ML, especially deep learning, has been facing stiff opposition, both in academia and industry, because of their black-box nature, lack of interpretability and generalizability which might produce nonphysical results \cite{faghmous2014theory, wagner2016theory, karpatne2017theory}. A perspective on machine learning for advancing fluid mechanics is available in a recent review article \cite{brenner2019perspective}.

Hybridization of both of the aforementioned approaches is therefore sought to maximize their pros and mitigate their cons \cite{reichstein2019deep}. Several research efforts have been devoted to achieving this hybridization, for example, in the form of closure modeling \cite{rahman2018hybrid, san2018neural, wan2018data, xie2018data, mohebujjaman2019physically}, accelerating simulations \cite{maulik2019accelerating}, and enforcing physical laws by tailoring loss functions \cite{raissi2019physics, pan2019physics, zhu2019physics}. In this paper, we address the problem of hidden physics, or unknown source terms by utilizing a neural network with the assumption that observed data is a manifestation of all interacting mechanisms and sources. In particular, a long-short term memory (LSTM) architecture, a variant of recurrent neural networks, is used to account for the unknown physics by learning a correction term representing the discrepancy between the physical model and actual observations. This is the part where intrusive ROM fails. Meanwhile, the generalizability of the model under different operating conditions is retained by employing intrusive (physics-based) ROM to represent the core dynamics. In other words, the underlying physics is divided into two parts, known part (core physics) modeled by intrusive approaches (e.g., Galerkin ROM) and unknown part (hidden physics) modeled by nonintrusive approaches (e.g., LSTM). A similar framework was presented in another study where we proposed a modular hybrid analysis and modeling (HAM) approach to account for hidden physics in parameterized systems \cite{pawar2019data}. The main difference between the two studies is the mechanism of adding a correction to the Galerkin projection model. In the previous study, the GP model is corrected dynamically with the LSTM model at each timestep and the corrected modal coefficients are used to predict the future state recursively. In the present study, we propose an evolve-then-correct (ETC) approach to account for the hidden physics where the GP and LSTM models are segregated. First, the GP model is used to evolve the initial state to the final time based on the known core dynamics. Then, a correction from the LSTM model is added to the time instant of interest. This means that the GP model evolves with uncorrected modal coefficients and the correction is enforced statically as a final post-processing step.

Relevant to our ETC approach, an evolve-then-filter approach has been introduced by Wells et al \cite{wells2017evolve} as ROM regularization. In their study, GP ROM is evolved for one timestep, after which a spatial filter is applied to filter the intermediate solution obtained in the evolve step. This filtering reduces the numerical oscillation of the flow variables (i.e., adds numerical stabilization to ROM). More recently, Gunzburger et al \cite{gunzburger2019evolve} proposed an evolve-filter-relax approach for uncertainty quantification of the time-dependent Navier-Stokes equations in convection dominated regimes. This is similar to the evolve-then-filter approach with the additional step of relaxation which averages the unfiltered and filtered flow variables to control the amount of numerical dissipation introduced by the filter. In the present paper, we put forth a nonintrusive memory embedding neural network model to take hidden physics into account for superparameterized systems. An illustration of the proposed framework is provided in solving fluid flow problems dynamically governed by the two-dimensional Navier-Stokes equations. We highlight that our ETC is modular in the sense that it does not require any change in the legacy codes and can be considered as an enabler for parametric model order reduction of partial differential equations.

\section{Evolve-then-correct approach}
\label{sec:etc}
In this study we consider the nonlinear dynamical system parameterized with a parameter $\mu$ which has the form

\begin{equation} \label{eq:dyn_system}
    \mathbf{u}_t(\mathbf{x},t;\mu, \kappa)=\mathbf{F}(\mathbf{x},t;\mathbf{u};\mu) + \mathbf{\Pi}(\mathbf{x},t;\mathbf{u};\mu,\kappa),
\end{equation}
where $\mathbf{u}$ is the state of the system, the subscript $t$ denotes the temporal derivative, $\mathbf{F}$ is the dynamical core of the system governing the known processes parameterized by $\mu$, and $\mathbf{\Pi}$ includes the unknown physics. The unknown physics encompasses deviation between the modeled and observed data resulting from several factors such as empirical parameterizations, and imperfect knowledge about the physical processes. The $\kappa$ in our study refers to the control parameter modeling the interaction between dynamical core of the system and hidden physics. 

We use the proper orthogonal decomposition (POD) to extract the dominant modes representing the above nonlinear dynamical system. We collect the data snapshots $\mathbf{u}_1, \mathbf{u}_2, \dots, \mathbf{u}_N$ $\in$ $\mathbb{R}^M$ at different time instances. In POD, we construct a set of orthonormal basis functions that optimally describes the field variable of the system. We form the rectangular matrix $\mathbf{A} \in \mathbb{R}^{M \times N}$ consisting of the the data snapshot $\mathbf{u}_n$ as its column. Then we use singular value decomposition (SVD) to compute the left and right singular vectors of the matrix $\mathbf{A}$. In matrix form the SVD can be written as 
\begin{equation}
    \mathbf{A} = \mathbf{W}\mathbf{\Sigma}\mathbf{V}^T = \sum_{k=1}^N\sigma_k \mathbf{w}_k \mathbf{v}_k^T,
\end{equation}
where $\mathbf{W}\in \mathbb{R}^{M \times N}$, $\mathbf{\Sigma} \in \mathbb{R}^{N \times N}$, and $\mathbf{V} \in \mathbb{R}^{N \times N}$. The $\mathbf{W}$ and $\mathbf{V}$ contains the left and right singular vectors which are identical to the eigenvectors of $\mathbf{A} \mathbf{A}^T$ and $\mathbf{A}^T \mathbf{A}$, respectively. Also, the square of singular vales are equal to the eigenvalues, i.e., $\lambda_k = \sigma_k^2$. The vectors $\mathbf{w_k}$ (also the eigenvectors of $\mathbf{A} \mathbf{A}^T$) are the POD basis functions and we denote them as $\phi_k$ in this text. The POD basis functions are orthonormal (i.e., $\langle \phi_i, \phi_j \rangle = \delta_{ij}$) and are computed in the optimal manner in the $L_2$ sense \cite{holmes2012turbulence, rowley2017model}. The full order model (FOM) for the dynamical system can be approximated using these POD basis functions as follow,
\begin{equation} \label{eq:rom_construction}
    \mathbf{u}(\mathbf{x}, t) = \sum_{k=1}^{R} a_k(t) \phi_k(\mathbf{x}),
\end{equation}
where $R$ of the number of retained basis functions such that $R << N$ and $a_k$ are the time dependent modal coefficients. The POD basis functions minimizes the mean-square error between the field variable and its truncated representation. Moreover, it also minimizes the number of basis functions required to describe the field variable for a given error \cite{1054286}. The number of retained modes is usually decided based on their energy content. Using these retained modes we can form the POD basis set $\mathbf{\Phi}=\{ \phi_k\}_{k=1}^{R}$ to build the ROM.  

POD is often complemented with Galerkin projection (GP) to reduce higher-dimensional partial differential equations (PDEs) into reduced-order ordinary differential equations (ODEs). To get GP equations, first we use a linear superposition of approximated field variable given by Equation~\ref{eq:rom_construction} into the governing equation of the physical system. Then we apply inner product of the resulting equation with the orthonormal basis function $\phi_k$. Therefore, we need the complete information about the governing equation of the physical system to form the GP equations that can describe the system accurately. We do not know about the hidden physics given by the source term $\mathbf{\Pi}$ in Equation~\ref{eq:dyn_system}. Therefore, we cannot derive a fully intrusive GP model for such dynamical system. 
\begin{figure}[htbp]
\centering
\mbox{
\subfigure{\includegraphics[width=0.45\textwidth]{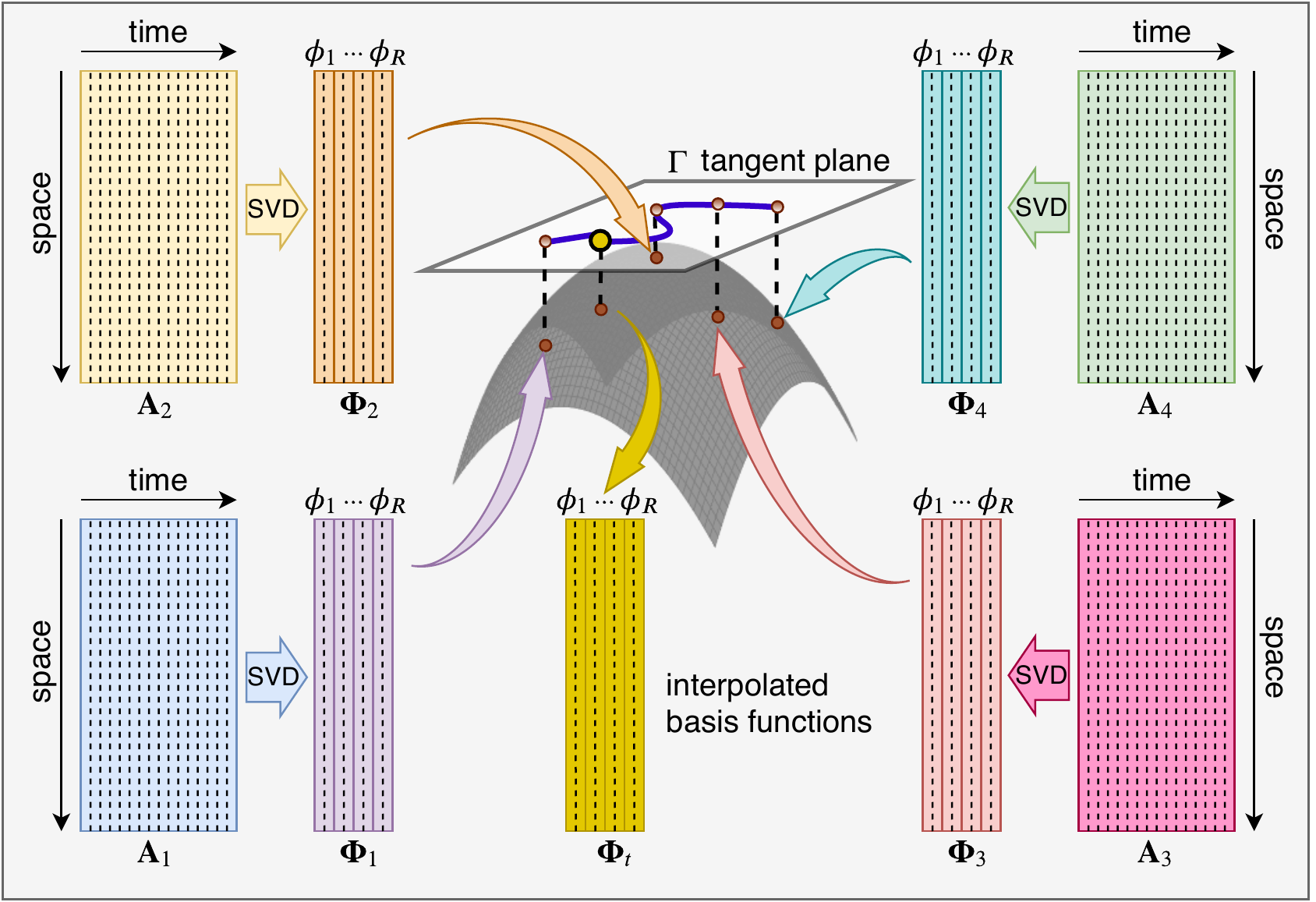}}
}
\caption{An illustration of the Grassmann manifold interpolation procedure to construct basis functions for a new testing condition. }
\label{fig:gm}
\end{figure}

In this study, we use machine learning algorithm to model this source term due to hidden physics and GP equations are derived based on the dynamical core of the system $\mathbf{F}$. The GP equations for the system with linear and nonlinear operator can be written as 
\begin{equation} \label{eq:gp0}
     \mathbf{\dot{a}} =  \mathbf{\mathfrak{L}} \mathbf{a} + \mathbf{a}^T \mathbf{\mathfrak{N}} \mathbf{a},
\end{equation}
or more explicitly,
\begin{equation} \label{eq:gp}
    \dfrac{\text{d}a_k}{\text{d}t} =  {\sum_{i=1}^{R} \mathfrak{L}^{i}_{k} a_i + \sum_{i=1}^{R} \sum_{j=1}^{R} \mathfrak{N}^{ij}_{k} a_i a_j} ,
\end{equation}
where $\mathfrak{L}$ and $\mathfrak{N}$ are the linear and nonlinear operator of the physical system. We limited our formulation in considering quadratic nonliterary without loss of generality by using $R$ modes. We use third-order Adams-Bashforth (AB3) to numerically integrate Equation~\ref{eq:gp}. In the discrete sence, the update formula can be written as 
\begin{equation}
    {a}_k^{(n+1)} = {a}_k^{(n)} + {\Delta t} \sum_{q=0}^{s} \beta_q G({a}_k^{(n-q)}),   
\end{equation}
where $s$ and $\beta_q$ are the constants corresponding to AB3 scheme, which are $s=2,~ \beta_0=23/12,~ \beta_1=-16/12,$ and $\beta_2=5/12$. We can obtain the true projection modal coefficients by simply projecting the field variable onto the basis functions and can be written as 
\begin{equation}
    \alpha_k^{(n)} = \langle \mathbf{u}(\mathbf{x}, t_n) , \phi_k \rangle,
\end{equation}
where the angle-parentheses refers to the Euclidean inner product defined as $\langle \mathbf{x} , \mathbf{y} \rangle = \mathbf{x}^T \mathbf{y} =  \sum_{i=1}^{M} x_iy_i$. The true projection modal coefficients includes the hidden physics and its interaction with the dynamical core of the system. The GP modal coefficients given by Equation~\ref{eq:gp} does not model this effect. We can then define the correction as 
\begin{equation} \label{eq:correction}
    \text{Correction} = \alpha_k^{(n)} - a_k^{(n)}.
\end{equation}
The above correction term can be learned with data-driven machine learning algorithm \cite{mou2019data, rahman2018hybrid}. We employ long-short term memory architecture \cite{hochreiter1997long}, a variant of recurrent neural network to learn this correction. LSTM are particularly suitable for time-series prediction, since they can use the information about the previous state of the system to predict the next state of the system. We train our LSTM neural network to learn the mapping from GP modal coefficients to the correction term, i.e., $\{a_1,\dots,a_R\}\in \mathbb{R}^R \rightarrow \{C_1,\dots,C_R\} \in \mathbb{R}^R$, where $C$ is the correction given by Equation~\ref{eq:correction}. We use three lookbacks to be consistent with AB3 scheme during training (please see the details in \cite{rahman2019non}). Since, GP modal coefficients are used as input features to the LSTM network, the parameter $\mu$ governing the system's behavior is taken implicitly into account. Once the model is trained, we can correct the GP modal coefficients with LSTM based correction to approximate true projection modal coefficients. 
\[
 \boxed{\text{ETC ROM} = \text{GP ROM} + \text{LSTM Correction}}
 \]
The ROM requires the basis functions which were obtained using POD. The dataset for the POD is generated using FOM. Running FOM for every ROM is opposite to the motivation behind using ROM. Hence, we should be able to generate the POD basis from existing POD basis sets using interpolation. The training data is generated for different values of parameters $\mu_1,\dots,\mu_P$. We compute the separate basis set $\mathbf{\Phi}_1,\dots,\mathbf{\Phi}_P$ for each of these parameter. We employ Grassmann manifold interpolation \cite{amsallem2008interpolation, zimmermann2018geometric} to compute the POD basis set for test parameter which is not included in the training set. The graphical illustration of the Grassmann manifold interpolation is provided in Figure~\ref{fig:gm} and the procedure is described in Algorithm~\ref{alg:gm}. 

\begin{breakablealgorithm}
  \caption{Grassmann manifold interpolation}
  \label{alg:gm}
  \begin{algorithmic}[1]
    \State Given a set of basis functions $\mathbf{\Phi}_1, \mathbf{\Phi}_2, \dots, \mathbf{\Phi}_P$ corresponding to the offline simulations (i.e., with mapping $S_1, S_2, \cdots, S_P$) paramaterized by $\mu_1, \mu_2, \dots, \mu_P$.
    \State Select a point $S_{0} \leftarrow S_i \in [S_1, \dots, S_P]$ corresponding to the basis function set $\mathbf{\Phi}_0 \leftarrow \mathbf{\Phi}_i \in [\mathbf{\Phi}_1, \dots, \mathbf{\Phi}_P]$ as the reference point.
    \State Map each point $S_i$ to a matrix $\mathbf{\Gamma}_i$ which represents the tangent space using logarithm map $\text{Log}_{S_0}$
    \begin{equation}
    (\mathbf{\Phi}_i-\mathbf{\Phi}_0 \mathbf{\Phi}_0^T \mathbf{\Phi}_i)(\mathbf{\Phi}_0^T\mathbf{\Phi}_i)^{-1} = \mathbf{U}_i\mathbf{\Sigma}_i \mathbf{V}_i^T,
    \end{equation}
    \begin{equation}
    \mathbf{\Gamma}_i =  \mathbf{U}_i \text{tan}^{-1}(\mathbf{\Sigma}_i) \mathbf{V}_i^T.
    \end{equation}
    \State Construct matrix $\mathbf{\Gamma}_t$ corresponding to the test parameter $\mu_t$ using the Lagrange interpolation of matrices $\mathbf{\Gamma}_i$, corresponding to $\mu_1,\dots,\mu_P$
    \begin{equation}
    \mathbf{\Gamma}_t = \sum_{i=1}^{P}\bigg( \prod_{\substack{j=1 \\ j\neq i}}^{P}\frac{\mu_t - \mu_j}{\mu_i - \mu_j}\bigg)\mathbf{\Gamma}_i.
    \end{equation}
    \State Compute the POD basis functions $\mathbf{\Phi_t}$ corresponding to the test parameter $\mu_t$ using the exponential map.
    \begin{equation}
    \mathbf{\Gamma}_t = \mathbf{U}_t \mathbf{\Sigma}_t \mathbf{V}_{t}^{T},
    \end{equation}
    \begin{equation}
    \mathbf{\Phi}_t = [\mathbf{\Phi}_0 \mathbf{V}_t \text{cos}(\mathbf{\Sigma}_t) + \mathbf{U}_t \text{sin}(\mathbf{\Sigma}_t)]\mathbf{V}_{t}^{T},
    \end{equation}
  where the trigonometric operators apply only to the diagonal elements.
  \end{algorithmic}
\end{breakablealgorithm}

\section{Numerical results}
\label{sec:results}

\begin{figure}[htbp]
\centering
\mbox{
\subfigure{\includegraphics[width=0.45\textwidth]{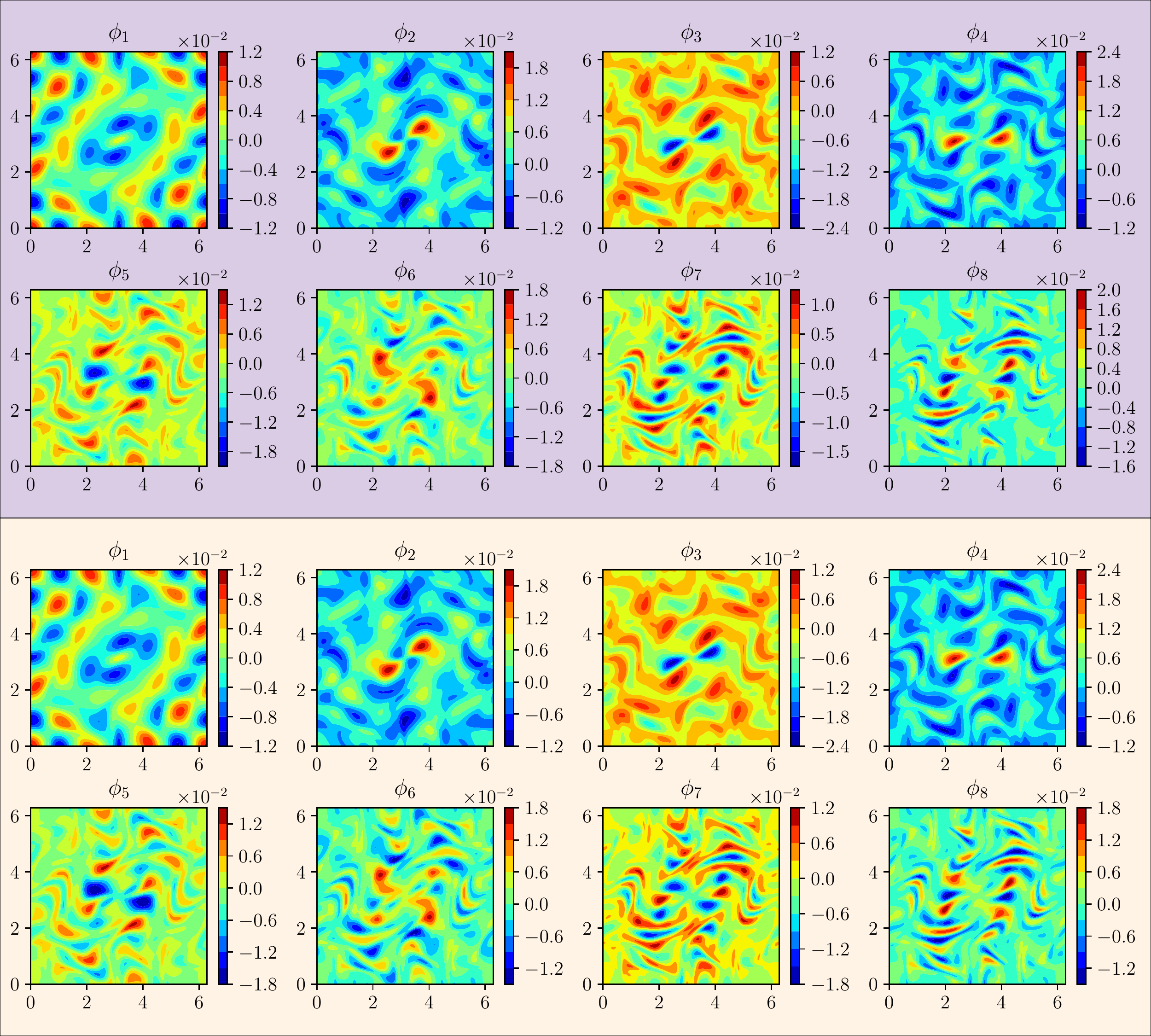}}
}
\caption{POD basis functions for $\text{Re}=1000$ with $\gamma=0.1$, true basis functions (top) and Grassmann interpolated basis functions (bottom).}
\label{fig:basis}
\end{figure}

We demonstrate the performance of evolve-then-correct (ETC) framework discussed in Section~\ref{sec:etc} for the two-dimensional Navier-Stokes equations. We use vorticity-streamfunction formulation of Navier-Strokes equations and it can be written as 
\begin{align} \label{eq:ns2d}
\dfrac{\partial \omega}{\partial t} + \dfrac{\partial \psi}{\partial y}\dfrac{\partial \omega}{\partial x} - \dfrac{\partial \psi}{\partial x} \dfrac{\partial \omega}{\partial y} &= \dfrac{1}{\text{Re}}\left( \dfrac{\partial^2 \omega}{\partial x^2} + \dfrac{\partial^2 \omega}{\partial y^2} \right) + \Pi,
\end{align}
\begin{align} \label{eq:ns2d-mass}
\dfrac{\partial^2 \psi}{\partial x^2} + \dfrac{\partial^2 \psi}{\partial y^2}=-\omega 
\end{align}
where $\omega$ is the vorticity defined as $\omega = \nabla \times \mathbf{u}$, $\mathbf{u} = [u,v]^T$ is the velocity vector, and $\psi$ is the streamfunction. We use vortex merger as the test example. In this test example, a pair of co-rotating vortices are separated from each other by some distance. These vortices induce fluid motion and strongly interact in the merging process. Vortex merger is extensively studied in the two-dimensional context as it explains the average inverse energy cascades and the direct enstrophy cascade observed in two-dimensional turbulence \cite{reinaud2005critical}. The initial condition for the vortex-merger test case is given as 
\begin{align}
    \omega(x, y, 0) = & \exp\left( -\pi \left[ (x-x_1)^2  + (y-y_1)^2 \right] \right) +  \nonumber \\ & \exp{\left( -\pi \left[ (x-x_2)^2 + (y-y_2)^2 \right] \right)},
\end{align}
where their centers are initially located at $(x_1,y_1) = (3\pi/4,\pi)$ and $(x_2,y_2) = (5\pi/4,\pi)$. We utilize an arbitrary array of Taylor-Green vortices as the source term that represents a perturbation field (i.e., referring to hidden physics). The source term in Equation~\ref{eq:ns2d} is given below
\begin{equation}
    \Pi = -F(t)\text{cos}(\eta x) \text{cos}(\eta y),
\end{equation}
where $F(t)= \gamma~ e^{-t/\text{Re}}$ and $\eta=3$. We use computational domain $(x,y) \in [0,2\pi]$ with periodic boundary conditions. We generate data snapshots for $\text{Re}=[200,400,600,800]$ with $256^2$ spatial grid and a time-step of 0.01 from time $t=0$ to $t=20$. We test the ETC framework for out-of-sample condition at $\text{Re}=1000$. The linear and nonlinear operators in GP equations for two-dimensional Navier-Stokes equation are 
\begin{align}
    \mathfrak{L}^{i}_{k} &= \bigg \langle \dfrac{1}{\text{Re}} \bigg(\dfrac{\partial^2 \phi_i^{\omega}}{\partial x^2} + \dfrac{\partial^2 \phi_i^{\omega}}{\partial y^2}\bigg),\phi_k^{\omega} \bigg \rangle, \\
    \mathfrak{N}^{ij}_{k} &= \bigg \langle -\bigg(\dfrac{\partial \phi_i^{\omega}}{\partial x} \dfrac{\partial \phi_j^{\psi}}{\partial y} -  \dfrac{\partial \phi_i^{\omega}}{\partial y} \dfrac{\partial \phi_j^{\psi}}{\partial x}\bigg),\phi_k^{\omega} \bigg \rangle,
\end{align}
where $\phi_k^{\omega}$ and $\phi_k^{\psi}$ refer to POD basis functions of the vorticity and streamfunction fields, respectively \cite{san2015stabilized}.
We retain 8 basis functions (i.e., $R=8$) as it captures more than 99.95\% energy for all Reynolds number included in training. We illustrate ETC framework for two different amplitudes of the source term, $\gamma=0.01$ and $\gamma=0.1$. 

\begin{figure}[htbp]
\centering
\mbox{
\subfigure{\includegraphics[width=0.45\textwidth]{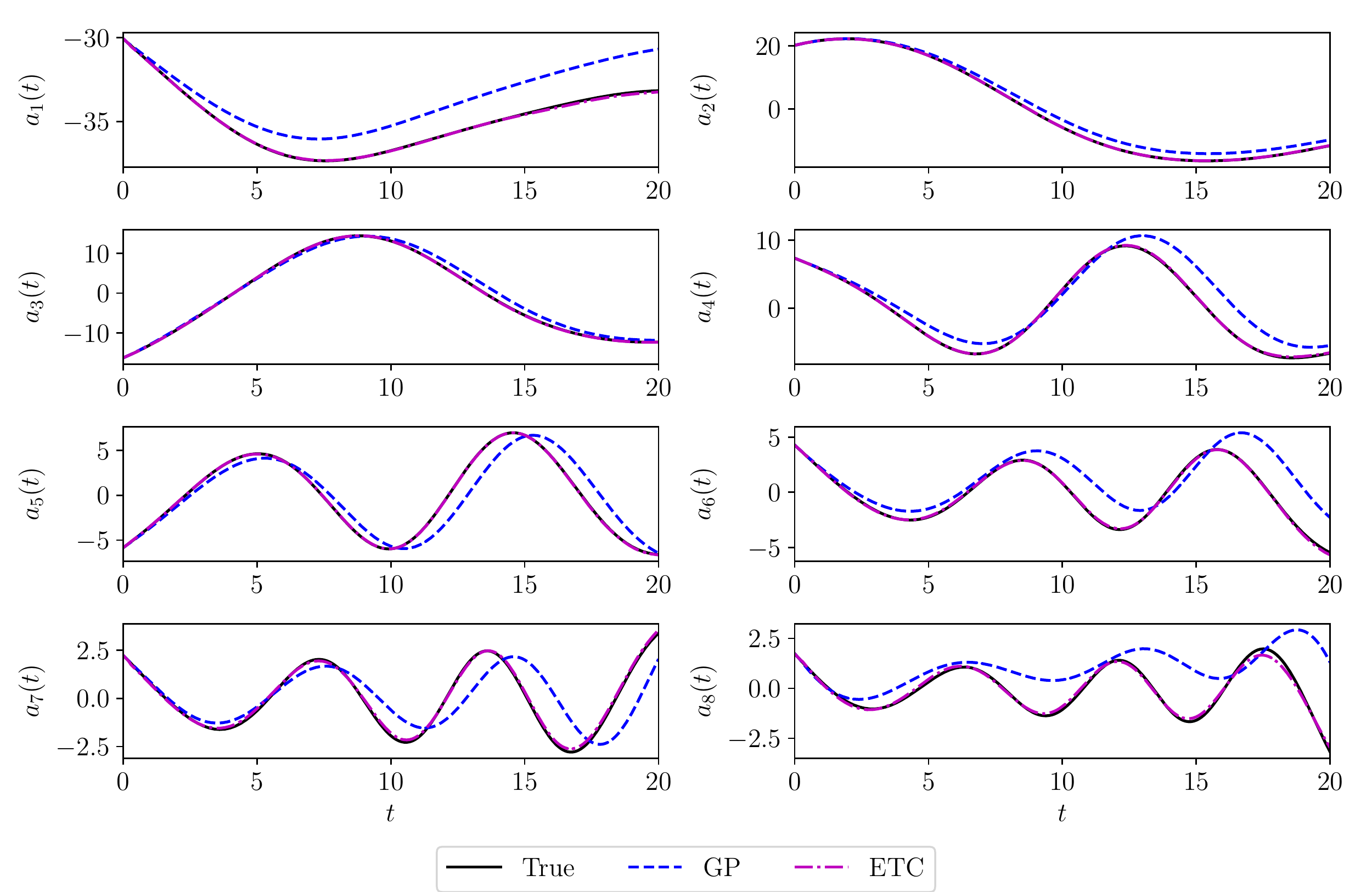}}
}
\caption{Evolution of vorticity modal coefficients at $\text{Re}=1000$ with $\gamma=0.01$. }
\label{fig:h2_s1_modes_1000}
\end{figure}

We illustrate the ETC approach for two different magnitudes of the source term, $\gamma=0.01$ and $\gamma=0.1$. We use $\text{Re}=800$ as the reference Reynolds number for both test cases. Figure~\ref{fig:basis} shows the true basis functions and Grassmann interpolated basis functions at $\text{Re}=1000$. The Grassmann manifold interpolation procedure can compute correct basis functions with very little deviation (especially for basis $\phi_7$ and $\phi_8$) from true basis functions. We train the LSTM network with two hidden layers and 80 cells. Our experiments with different hyperparameters show that the results are not highly sensitive to the choice of hyperparameters. 

\begin{figure}[htbp]
\centering
\mbox{
\subfigure{\includegraphics[width=0.45\textwidth]{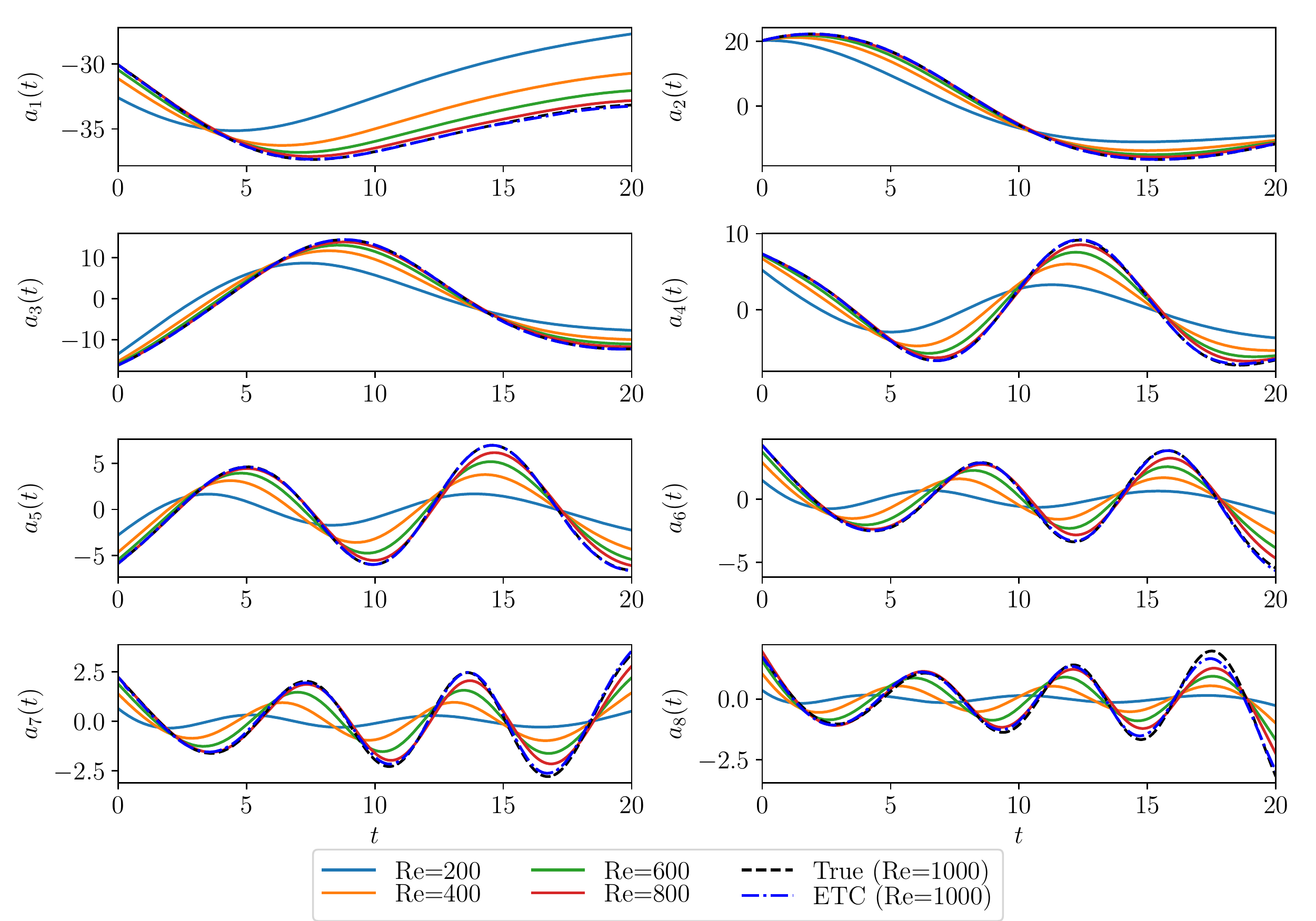}}
}
\caption{Evolution of vorticity modal coefficients at different training Reynolds numbers and test Reynolds number ($\text{Re}=1000$) with $\gamma=0.01$. }
\label{fig:h2_s1_allmodes_1000}
\end{figure}

Figure~\ref{fig:h2_s1_modes_1000} shows the evolution of vorticity modal coefficients for $\gamma=0.01$. The GP modal coefficients are different from the true projection modal coefficients even though the 8 modes are capturing more than 99.95\% of the energy. This difference is due to the source term not modeled by GP equations. In the ETC approach, we correct the GP modal coefficients and we get an excellent agreement with the true modal coefficients after correction. The vorticity modal coefficients obtained with the ETC approach is almost the same as the true projection modal coefficients. 

\begin{figure}[htbp]
\centering
\mbox{
\subfigure{\includegraphics[width=0.45\textwidth]{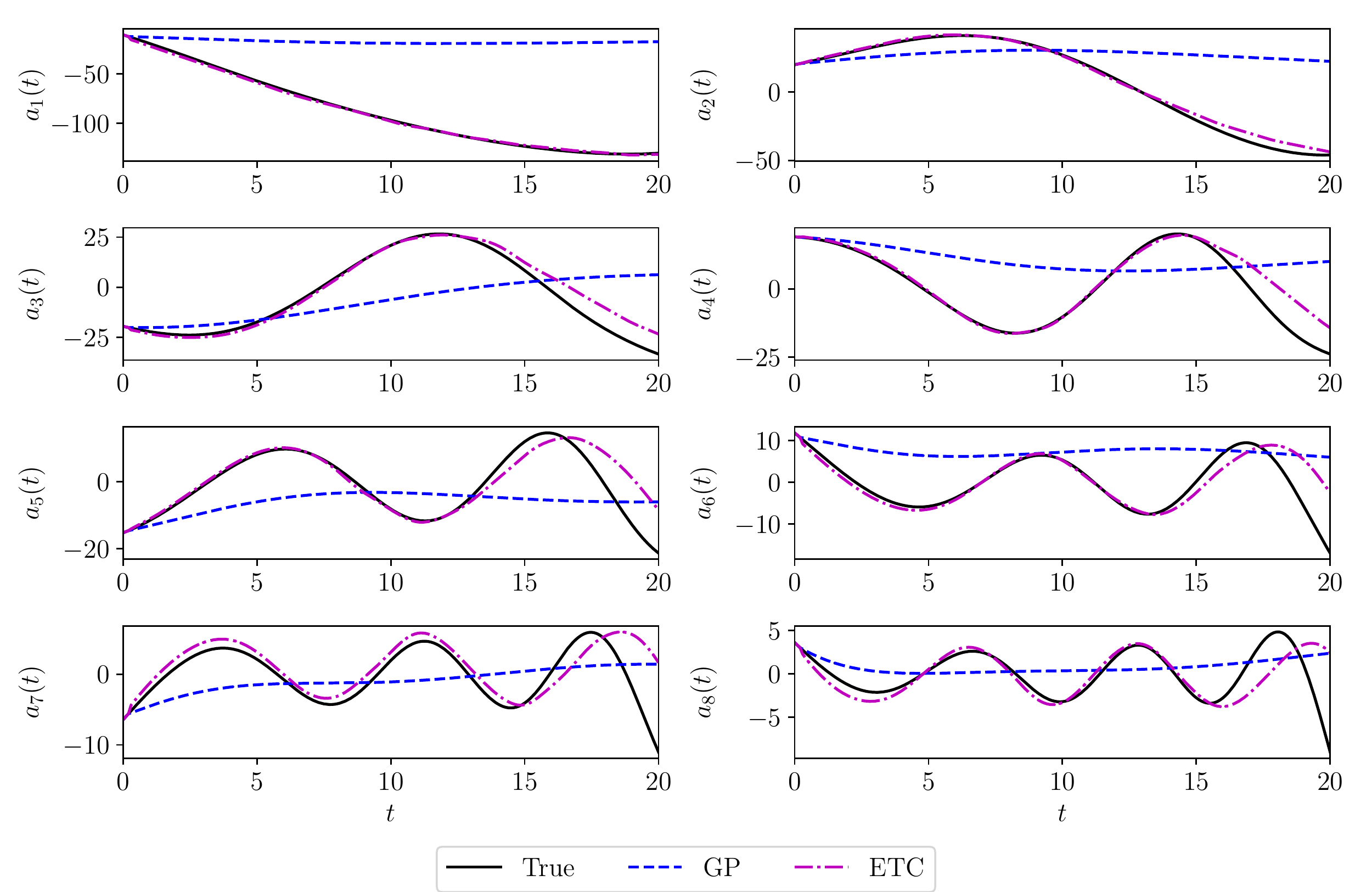}}
}
\caption{Evolution of vorticity modal coefficients at $\text{Re}=1000$ with $\gamma=0.1$. }
\label{fig:h2_s2_modes_1000}
\end{figure}

To show the difference between vorticity modal coefficients at all train Reynolds number and the test Reynolds number, we plot time series of modal coefficients for all Reynolds number in the same plot as reported in Figure~\ref{fig:h2_s1_allmodes_1000}. We can observe that the evolution of vorticity modal coefficients for the test Reynolds number is different from train Reynolds number and the ETC approach is able to produce the correct trajectory of modal coefficients at test Reynolds number. This also shows that the LSTM has not simply memorized the correction from train Reynolds number, but it has learned the dependence of modal coefficients on the parameter (through input features) governing the physical system.

\begin{figure}[htbp]
\centering
\mbox{
\subfigure{\includegraphics[width=0.45\textwidth]{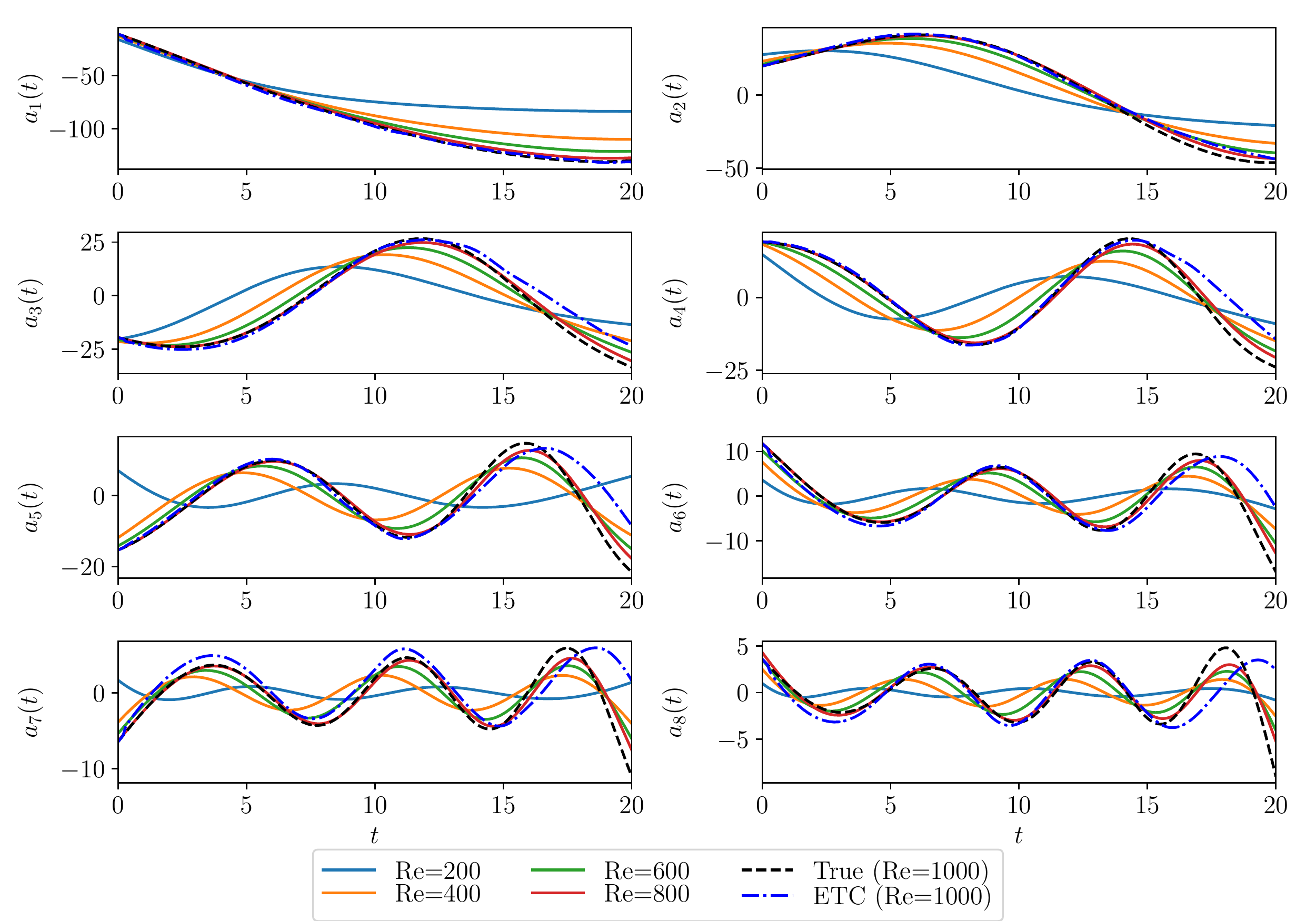}}
}
\caption{Evolution of vorticity modal coefficients at different training Reynolds numbers and test Reynolds number ($\text{Re}=1000$) with $\gamma=0.1$. }
\label{fig:h2_s2_allmodes_1000}
\end{figure}

Figure~\ref{fig:h2_s2_modes_1000} displays the evolution of modal coefficients for $\gamma=0.1$. We see that there is a large deviation between GP modal coefficients and true projection modal coefficients due to a large magnitude of the source term. The ETC approach can correct GP modal coefficients and produce the modal coefficients trajectories close to the true projection. Since the source term is very large and has the magnitude the same as the main field, we do not get the same level of accuracy as the test case with $\gamma=0.01$, especially near the end time (i.e., $t=15-20$). Figure~\ref{fig:h2_s2_allmodes_1000} presents the vorticity modal coefficients at all train Reynolds number and the test Reynolds number.

\begin{figure*}[htbp]
\centering
\mbox{
\subfigure{\includegraphics[width=0.95\textwidth]{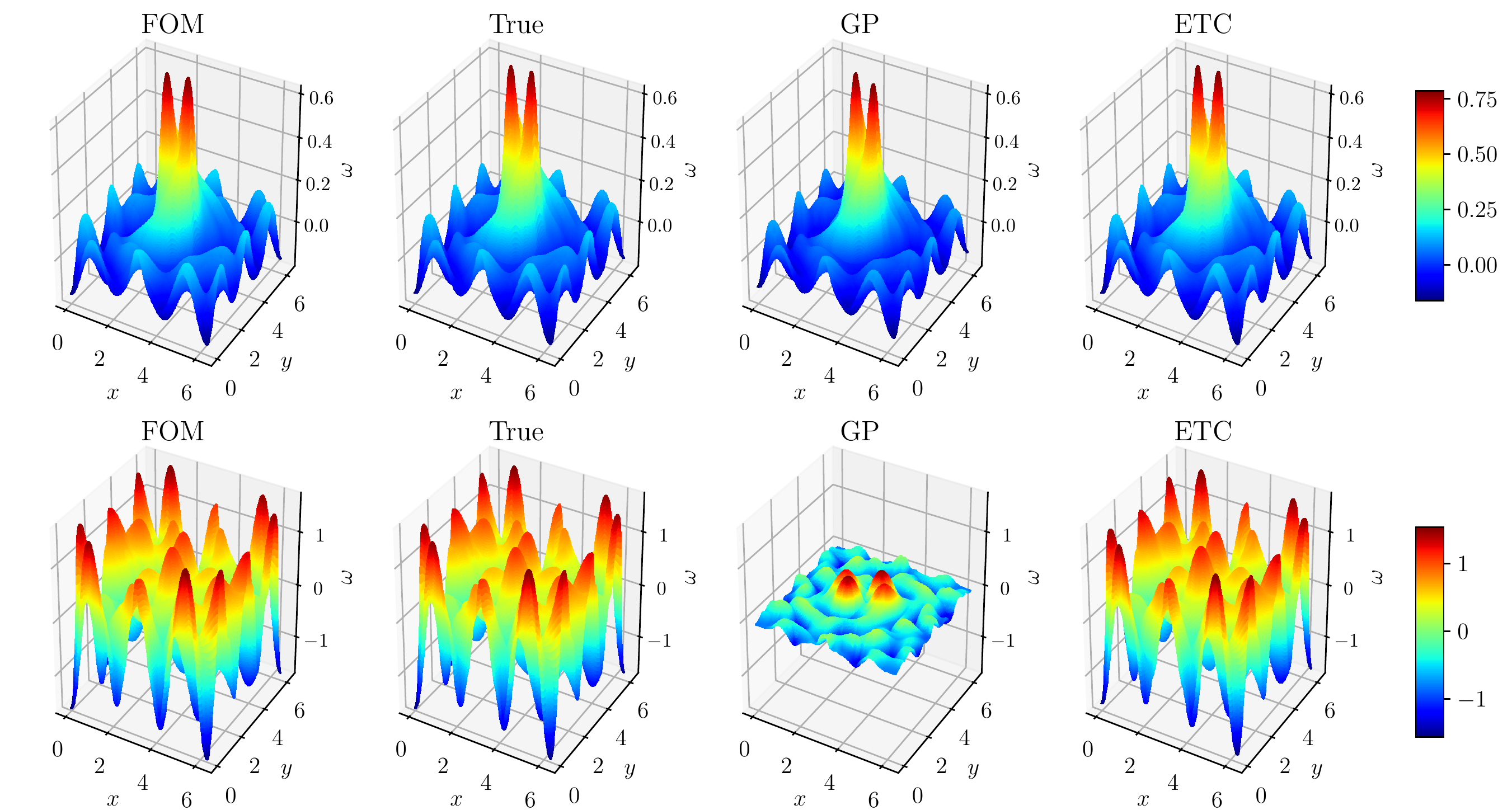}}
}
\caption{Vorticity field at the final time $t=20$ for $\text{Re}=1000$, $\gamma=0.01$ (top) and $\gamma=0.1$ (bottom). }
\label{fig:vort_field}
\end{figure*}

Finally, Figure~\ref{fig:vort_field} shows the three-dimensional plot of the reconstructed vorticity field computed with FOM, true projection, GP, and ETC modal coefficients. We do not see a significant difference between the three-dimensional vorticity field computed with GP, and ETC modal coefficients for $\gamma=0.01$. However, the orientation of two vortices is not correctly captured by GP modal coefficients in comparison to the true projection. The ETC approach correctly predicts the orientation of two vortices close to the true projection. There is a large difference between the reconstructed vorticity field for GP and true projection at $\gamma=0.1$. The magnitude of the reconstructed vorticity field with GP modal coefficients is very small compared to the true projection modal coefficients and it is not able to produce the vorticity field due to the Taylor-Green vortices. The ETC approach correctly predicts the vorticity field with sufficient accuracy and also produces the vorticity field due to Taylor-Green vortices. The correct prediction of the vorticity field for large source term illustrates the advantage of the ETC approach for physical systems where there is a discrepancy between the modeled and observed data due to modeling assumptions, imperfect parameterizations, and insufficient knowledge about the physical system.      

\section{Concluding remarks}
\label{sec:con}
We present an evolve-then-correct approach for reduced order modeling of parameterized systems with hidden information. This hidden information can be attributed to unknown source term, imperfect parameterizations, or incorrect modeling assumptions. We correct the physics-based model (GP model in this study) with the hidden information recovered by ML algorithms (LSTM neural network in our framework).

We demonstrate the performance of the ETC approach for two-dimensional Navier-Stokes equations used to simulate the merging of co-rotating vortices. Our numerical experiments with two different magnitudes of source term yield highly accurate solutions close to true projection results. We also show that the LSTM algorithm has effectively learned the relationship between the unknown source term and the parameter governing the system's behavior through GP modal coefficients as input features. 

We generate the synthetic source term using an array of Taylor-Green vortices in our numerical experiments and the results of the ETC approach are promising. However, there are several research directions that we plan to pursue in the future to better understand the capability and limitations of the ETC approach. One of the future research directions is to test the ETC approach with more realistic noisy data obtained from sensors. \\

\section*{Acknowledgement}
This material is based upon work supported by the U.S. Department of Energy, Office of Science, Office of Advanced Scientific Computing Research under Award Number DE-SC0019290. O.S. gratefully acknowledges their support. 

Disclaimer. This report was prepared as an account of work sponsored by an agency of the United States Government. Neither the United States Government nor any agency thereof, nor any of their employees, makes any warranty, express or implied, or assumes any legal liability or responsibility for the accuracy, completeness, or usefulness of any information, apparatus, product, or process disclosed, or represents that its use would not infringe privately owned rights. Reference herein to any specific commercial product, process, or service by trade name, trademark, manufacturer, or otherwise does not necessarily constitute or imply its endorsement, recommendation, or favoring by the United States Government or any agency thereof. The views and opinions of authors expressed herein do not necessarily state or reflect those of the United States Government or any agency thereof.


\bibliography{reference}
\bibliographystyle{apsrevlong}

\end{document}